\documentclass[11pt,a4paper, table]{article}
\usepackage[hyperref]{emnlp2018}
\usepackage{times}
\usepackage{latexsym}
\usepackage{url}
\usepackage{xcolor}
\usepackage{tabularx}
\aclfinalcopy % Uncomment this line for the final submission
\title{End-to-End Retrieval in Continuous Space}

\author{Daniel Gillick, Alessandro Presta, Gaurav Singh Tomar \\
  Google AI \\
  {\tt \{dgillick, apresta, gtomar\}@google.com}  \\}

\date{}

\begin{document}
\maketitle
\begin{abstract}
Most text-based information retrieval (IR) systems index objects by words or phrases. These discrete systems have been augmented by models that use embeddings to measure similarity in continuous space. But continuous-space models are typically used just to re-rank the top candidates. We consider the problem of end-to-end continuous retrieval, where standard approximate nearest neighbor (ANN) search replaces the usual discrete inverted index, and rely entirely on distances between learned embeddings. By training simple models specifically for retrieval, with an appropriate model architecture, we improve on a discrete baseline by 8\% and 26\% (MAP) on two similar-question retrieval tasks. We also discuss the problem of evaluation for retrieval systems, and show how to modify existing pairwise similarity datasets for this purpose.
\end{abstract}

\section{Introduction}
\label{sec:introduction}
Nearly 30 years ago, Deerwester et al.~\shortcite{deerwester1990indexing} described the shortcomings of the standard retrieval systems that are still widely used today: "The problem is that users want to retrieve on the basis of conceptual content, and individual words provide unreliable evidence about the conceptual topic or meaning of a document." As a solution, they introduced Latent Semantic Indexing, using Singular Value Decomposition over word co-occurrences to \textit{encode} (or \textit{embed}) a piece of text as a dense low-dimensional vector rather than a sparse high-dimensional vector of word indicators. This work opened the field of representation learning \cite{bengio2013representation}, but did not address the issue of efficient retrieval from the learned space. We'll call the overall task -- constructing dense representations and retrieving neighbors -- \textit{continuous retrieval} by way of contrast with \textit{discrete retrieval} that uses an inverted index to leverage sparse representations. In principle, continuous retrieval has clear benefits: improved recall (unconstrained by specific word choice), more granular similarity scoring, learned relationships between query and candidates, and the possibility of retrieval across modalities.

%More recent work has made continuous retrieval feasible by employing Locality Sensitive Hashing \cite{indyk1998approximate,gionis1999similarity} to find approximate neighbors efficiently. While these methods have been useful for detecting duplicates and plagiarism \cite{stein2007principles}, standard search engines still rely on discrete retrieval (see Lucene\footnote{https://lucene.apache.org}, for example).

However, models for learning text representations have found application in IR by re-ranking the top candidates proposed by a discrete retrieval system \cite{huang2013learning,shen2014latent,palangi2016deep,dos2015learning,lei2016semi}. To the best of our knowledge, there have been no previous comparisons of end-to-end retrieval systems \cite{onal2017neural}. A model intended for re-ranking differs from a model intended for retrieval in two important ways. First, a re-ranking model has access to the raw representations of both query and candidate and can thus learn complex interactions \cite{parikh2016decomposable,gong2017natural}, whereas a retrieval model must encode queries and candidates independently to allow for fast neighbor look-up. Second, re-rankers can focus modeling power on the boundary encoderscases proposed by the discrete retrieval systems, while retrieval models must also perform well with random pairs.

%\begin{itemize}
%\item The model will have access to the raw representation of both query and candidates. This allows the use of powerful “interaction” features that compare parts of both query and candidate; it also allows for arbitrarily complex learned functions of the two inputs, including attention mechanisms over inputs in both query and candidate.
%\item The candidates retrieved by a discrete system will already be somewhat relevant to the query (e.g. by means of word overlap). Thus, the comparison model can focus on discriminating the boundary cases, without concern for the full distribution of query/candidate pairs.
%\end{itemize}

%\begin{itemize}
%\item The query and candidate need to be encoded independently, and the two representations need to be combined with a simple metric (e.g. cosine similarity) that allows for fast neighbor lookup. Complex interaction functions or features are not allowed.
%\item The learned model needs to perform well both at the boundary and with random pairs. Otherwise, the true positives will be overwhelmed by false positives as the candidate set size increases.
%\end{itemize}

The primary goal of this paper is to show that using standard ANN search, simple models trained for the purpose of continuous retrieval can substantially outperform discrete retrieval systems. We show evidence for choosing a negative sampling method which we call \textit{in-batch sampled softmax}, and evaluate a variety of baselines and trained models on two pairwise datasets that we modify for the purpose of retrieval evaluation.

%This paper makes the following contributions. First, we introduce a modular \textit{dual encoder} architecture for multi-task representation learning where the input consists of sets of pairwise positive examples. Second, we show that training dual encoders with a specific kind of negative sampling, which we call \textit{batch softmax}, is both efficient and effective for continuous retrieval. Third, because the retrieval task is inherently different from the more frequently evaluated similarity task, we show how existing pairwise datasets can be used to build retrieval datasets. Finally, we show a series of experiments in which models trained for continuous retrieval outperform a variety of baselines.

\section{Dual Encoders}
\label{sec:architecture}

%Data with explicit labels are hard to find and expensive to collect. But the world is filled with pairs of items, connected by all sorts of implicit relationships. Language models and Word2Vec embeddings \cite{mikolov2013distributed} take advantage of such positive pairs (current word and its context). Crucially, negative examples are implied: while there are a number of words that could reasonably fit with some context, a random word, on average, will be a poor substitute for the observed word.

%Encoder/decoder architectures leverage such pairwise data for translation \cite{cho2014learning,sutskever2014sequence} and summarization \cite{rush2015neural} (again, words not observed are negative examples at each output position). Here, we are interested in training models for computing similarities between pairs of objects encoded in the same space. For retrieval, a large set of candidate items is encoded off-line so that we can efficiently retrieve the neighbors of a query, encoded on-line in the same space.

Neural network models for learning distance functions date back to early work on signature verification \cite{bromley1994signature}, later extended to face verification \cite{chopra2005learning}. This work and its descendants \cite[etc.]{yih2011learning,hu2014convolutional} refer to the models as \textit{siamese networks} because two similar objects are encoded by two copies of the same network (all parameters are shared). The Wsabie model \cite{weston2010large}, intended for classification with large label sets, learns embeddings for the inputs and outputs separately. The StarSpace model \cite{wu2017starspace} extends the idea of learned embeddings to more data types. More generally, we refer to the class of models in which pairs of items are encoded in a shared space, as \textit{Dual Encoders}. This is a modular architecture with the following components:

\begin{description}
\small
\item \textbf{Encoder}: An encoder is any learnable function $f(X)$ that takes an item $X$ as input and returns a $d$-dimensional real-valued encoding vector. Here, we focus on neural network functions $f$.
\item \textbf{Similarity Function}: A similarity function $sim(E_1, E_2)$ takes two encodings of the same dimension, and outputs a score in $[0, 1]$. Similarity functions can be arbitrarily complex, including neural networks that learn interactions between encodings, but to enable nearest neighbor search, we use cosine similarity, the standard for retrieval \cite{manning2008introduction}.
\item \textbf{Dual Encoder}: A dual encoder has the form $g(X_1,X_2) = sim(f_1(X_1), f_2(X_2))$ where $f_1,f_2$ are two possibly identical encoders. We additionally apply a learned affine transform, $\alpha g(\cdot, \cdot) + \beta$, which scales the similarity so it can be treated as a logit during training.
%\item \textbf{Multi-task Dual Encoder}: In a multi-task setup, instances of the same encoder may appear in multiple dual encoders trained on different datasets. The parameters are tied among those instances, so that they are shared by the different tasks. In principle, any parameter sharing scheme among encoders is allowed. For example, we share embeddings across all text encoders even if the encoded objects are different.
\end{description}

Note that while we train dual encoders for each pairwise dataset, including scaling parameters $\alpha,\beta$, retrieval requires only the individual trained encoders: the candidate items are encoded by the candidate encoder and indexed off-line; at inference time, the query is encoded by the query encoder and neighbors are retrieved from the candidate space according to cosine distance.

In our experiments, we train a very simple form of dual encoder for similar question retrieval. Much like the Paragram-Phrase setup \cite{wieting2015towards}, we use a single question encoder that represents the input with an average over word embeddings. Thus, the question encoder parameters are just the set of learned embeddings.

Some of our experiments use a multi-task setup, with up to 3 tasks. While there is a separate dual encoder for each task, they all share the same question encoder, so only the scaling parameters are task-specific. In multi-task training, we compute a task-specific loss, then take a weighted average to produce the overall loss; the weights are uniform in all experiments.

\subsection{Loss functions}
\label{ssec:losses}

Much of the relevant prior work on representation learning has focused on pairwise similarity \cite{hu2014convolutional,wieting2015towards,arora2017asimple,conneau2017supervised}, sometimes with the goal of re-ranking retrieval candidates.

If the training data consists of positive and negative example pairs, it is standard to minimize the logistic (cross-entropy) loss between true labels and model predictions.

But often, training data consists just of positive pairs. In the Word2Vec setting \cite{mikolov2013distributed} or in Language Model training \cite{jozefowicz2016exploring}, the negative examples are implied: while there are a number of words that could reasonably fit with some context, a random word, on average, will be a poor substitute for the observed word. These models are trained with a softmax loss, where the negatives are all non-observed words in the vocabulary. For efficiency, the denominator can be approximated with a sample from the vocabulary. In the more general dual encoder case, though, the set of negative examples may not be enumerable. Indeed, if both inputs are sentences (or questions), negative sampling is a necessary approximation.

We consider  a few different loss functions (in addition to the standard cross-entropy loss for binary-valued labels), each of which implies a different negative sampling strategy. All the strategies make use of items in the batch as a source of random negatives. A batch includes $B$ positive pairs of items which have been encoded by their respective encoders. We apply the similarity function to all pairs $(E_1^i, E_2^j)$ to form a similarity matrix $M$ where the diagonal contains positive examples and the off-diagonal contains random negative examples.

\begin{description}
\small
\item \textbf{In-batch Cross-Entropy} We form a cross-entropy loss term for each element in $M$, with positives on the diagonal and negatives on the off-diagonal, and return the average.
\item \textbf{In-batch Sampled Softmax} We form a softmax loss term for each row in $M$, where row $i$ has a positive label on column $i$ (corresponding to the diagonal), and return the average. This was suggested by Henderson et al. \shortcite{henderson2017efficient}.
\item \textbf{In-batch Triplet} We form a triplet loss term for each row in $M$ that maximizes the margin between the positive element and the highest scoring negative element in the row: $\max(0, \delta - s^+ + s^-)$, where $\delta=0.5$. This is most similar to the loss used by Wieting et al. \shortcite{wieting2015towards}.
\end{description}

\subsection{Training}
\label{ssec:training}

We train all our models using mini-batch Gradient Descent with the Momentum optimizer and a fixed learning rate of $0.01$. Unless otherwise noted, the batch size is $1000$ and the loss is in-batch sampled softmax. We use a lowercased unigram vocabulary and 300-dimensional embeddings, initialized randomly. We use no explicit regularization (like dropout), but rely on early stopping (based on tuning set evaluation) to avoid over-fitting. In-batch precision@1 (accuracy computed over each row of the similarity matrix $M$), averaged over the rows in  $M$, is our tuning metric, since this is a reasonable proxy for precision@1 computed over the full set of candidates, which in turn represents retrieval performance.

\section{Experimental Setup}
\label{sec:setup}

\subsection{Evaluating end-to-end retrieval}
\label{ssec:end2end}

%Relevant prior work on representation learning for text has focused on one of two tasks:
%\begin{description}
%\item \textbf{Pairwise similarity}: Given a pair of text objects (two queries, a query and a document, a query and an answer, etc.), predict whether the two elements are similar. If the ground truth is a binary label, metrics such as AUC, F1, and accuracy are computed. If the degree of similarity is represented by a range of scores, Pearson or Spearman correlation are generally used.
%\item \textbf{Re-ranking}: Given a query and a limited set of candidate results, predict which of the candidates are relevant to the query. When the relevance information is binary (a candidate is either “relevant” or “not relevant”), metrics such as MAP, MRR, Precision at K, and less commonly bpref are computed. For scored relevance judgments, Normalized Discounted Cumulative Gain (NDCG) has been used.
%\end{description}

Neither pairwise similarity tasks nor re-ranking tasks
are useful for evaluating end-to-end retrieval: the pairs of items are usually sourced using some heuristic or existing retrieval system. The resulting test data distribution is biased towards pairs selected by that system. Such test sets may fail to discriminate among models that have drastically different performance on random pairs, and it is particularly important that retrieval models be robust to all sorts of noisy candidates.

An offline retrieval task consists of (1) a set of test queries, (2) a set of candidate items (sufficiently large so as to be realistic), and (3) a set of (query, candidate) pairs labeled with relevance judgments. However, for any reasonable size candidate set, it's infeasible to have all pairs annotated by a human. As a result, all retrieval tasks are necessarily incomplete: only a small subset of relevant candidates are labeled, so we assume that all unlabeled candidates are not relevant. This issue is discussed at length by Buckley and Voorhees \shortcite{buckley2004retrieval}, who show that the Mean Average Precision (MAP) metric computed on an incomplete evaluation set correlates reasonably well with the MAP metric computed on a (significantly more) complete version of that evaluation set.

Computing full MAP on such a dataset can be computationally expensive (for each query, all candidates need to be scored). Instead, we only consider the top K results and compute MAP@K based on the following definition:

\[ MAP@K = \sum_{q_i \in Q_i} \frac{1}{R_i} \sum_{j=1}^{k} p_{i}^j r_{i}^{j} \]

where $Q_i$ is the set of test queries, $R_i$ is the number of known relevant candidates for $Q_i$, $p_{i}^j$ is precision@j for $q_i$, and $r_{i}^{j}$ is 1 if the j\textsuperscript{th} result is relevant to $q_i$, 0 otherwise.

\subsection{Approximate nearest neighbor search}
\label{ssec:approximate_search}
While the problem of nearest neighbor search \cite{indyk1998approximate,gionis1999similarity} is central to continuous retrieval, we're glossing over it here for two reasons. First, a simple quantization method \cite{guo2016quantization} works quite well for the tasks we consider; second, since we are more interested in analyzing modeling issues, we use exhaustive search to avoid any confounding effects linked to the choice of approximate search algorithm. Moreover, we found that approximate search is nearly as accurate as exhaustive search in our retrieval tasks: MAP@100 for approximate search declined no more than 0.4\% even as we increased the candidate set size from 20k up to 1M.

\subsection{Constructing retrieval tasks}
\label{ssec:construction}
We use a simple approach to turn a conventional similarity scoring or ranking task into an incomplete retrieval task. Given a test set with labeled pairs, we first build the graph induced by positive pairs. Next, we compute the transitive closure of the graph, which may yield additional positive pairs. Now, each element of a positive pair is considered a test query, and its neighbors in the transitive closure graph are the known positive results for that query. Finally, the set of candidates consists of all items found in the test set (either in a positive or a negative pair).

We apply this method to the Quora question pairs dataset\footnote{https://data.quora.com/First-Quora-Dataset-Release-Question-Pairs} and the AskUbuntu dataset\footnote{https://github.com/taolei87/askubuntu} \cite{dos2015learning,lei2016semi} to produce new retrieval evaluation sets. We use only the question titles in the AskUbuntu data, and leave the more complex problem of modeling (often much longer) question bodies to future work. In our experiments, we apply our trained encoder models to all pairs of (query, candidate) and evaluate MAP@100 on the resulting scores. While this means that our results are not comparable to previous reported work using these pairwise datasets, we provide results from a variety of baseline systems.

The AskUbuntu training set includes just positive pairs, so negative sampling is required. However, the Quora training set includes positive and negative examples (in roughly a 2:1 ratio). This allows us to compare standard cross-entropy loss with our negative sampling strategies.

\begin{table}[t]
\small
\setlength{\tabcolsep}{4pt}
\centering
\label{counts}
\begin{tabular}{|c|cc|}
\hline
 & \textbf{Quora} & \textbf{AskUbuntu} \\
\hline
Positive training pairs & 139306 & 13010 \\
Test queries & 9218 & 1224 \\
Candidates & 19081 & 4023 \\
Relevant candidates/query & 2.55 & 11.73 \\
\hline
\end{tabular}
\caption{End-to-end retrieval task statistics.}
\end{table}

Since we are interested in a training setting where a single model works well for both tasks, we also experiment with the Paralex dataset\footnote{http://knowitall.cs.washington.edu/paralex}, 18 million question-paraphrase pairs scraped from WikiAnswers.

\subsection{Baselines}

To facilitate meaningful comparison, we start with a few common baselines. First, because each candidate set includes all the test queries, an "identity" baseline simply retrieves the exact test query as the only matched candidate. Second, we use TFIDF and the BM25 algorithm \cite{robertson2009probabilistic} for discrete retrieval, standard baselines for retrieval comparisons \cite{hoogeveen2015cqadupstack}.

We also compare a variety of averaged word embeddings baselines, starting with uniform averaging of 300-dimensional pretrained word2vec embeddings. Next, following Arora et al. \shortcite{arora2017asimple}, we take a weighted average of pretrained embeddings using Inverse Document Frequency (IDF)\footnote{We found no advantage by using the SIF weighting or PCA subtraction proposed by Arora et al.}, and try 3 different settings for pre-training: standard word2vec, word2vec trained with the Paralex dataset (closer to the question domain), and Glove \cite{pennington2014glove} trained from Web (Common Crawl) data. We also try embedding each question using a 300-dimensional Skip-Thought model \cite{kiros2015skip}.

Note that in all cases, the score for a query-candidate pair is computed using cosine distance between the respective encodings.

\section{Analysis of Results}
\label{sec:results}

\begin{table}[t]
\small
\setlength{\tabcolsep}{4pt}
\centering
\label{fullresults}
\begin{tabular}{|lc|ccc|}
\hline
\textbf{Model} & \textbf{Training data} & \textbf{Quora} & \textbf{AskU} & \textbf{AVG} \\
\hline
Identity & - & 45.9 & 14.4 & 30.2 \\
TFIDF & - & 77.2 &	35.6 & 56.4 \\
Okapi BM25 & - & 83.7 & 36.5 & 60.1 \\
\hline
Avg-word2vec & News & 78.4 & 28.4 & 53.4 \\
IDF-word2vec & News & 85.4 & 33.1 & 59.3 \\
IDF-GloVe & Web & 85.2 & 33.4 & 59.3 \\
IDF-word2vec & Paralex & 86.0 & 33.5 & 59.8 \\
Skip-Thought & Books & 73.3 & 19.6 & 46.4 \\
\hline
Dual Encoder & Paralex (P) & 87.6 & 37.3 & 62.4 \\
Dual Encoder & Quora (Q) & 90.4 & 35.8 & 63.1 \\
Dual Encoder & AskUbuntu (A) & 84.5 & 45.9 & 65.2 \\
Dual Encoder & Q + A & 88.3 & 42.2 & 65.2 \\
Dual Encoder & P + Q & \textbf{90.5} & 37.3 & 63.9 \\
Dual Encoder & P + A & 87.5 & \textbf{46.0} & 66.7 \\
Dual Encoder & P + Q + A & 89.9 & 45.5 & \textbf{67.7} \\
\hline
\end{tabular}
\caption{MAP@100 retrieval results.}
\label{t:mapresults}
\end{table}

Table \ref{t:mapresults} shows MAP@100 results on the Quora and AskUbuntu retrieval tasks. First, we observe that while IDF-weighting the pretrained embeddings is useful, this is still not clearly better than the BM25 baseline. We show this is not a domain issue by training word2vec directly with Paralex data. However, the dual encoder trained with Paralex data is significantly better, and now improves on BM25 on both evaluations. Next, we are able to improve results quite a bit more by using in-domain training data. And finally, we get the best overall results by training a single multi-task dual encoder that combines data from all three tasks (note that we train the Paralex-only dual encoder to convergence before adding the multi-task loss).

In Section \ref{ssec:losses}, we enumerated a number of loss functions using different negative sampling strategies. Most importantly, we found that training a Quora-only model with standard cross-entropy (using the provided positive and negative training examples) was substantially worse than training with any of the negative sampling strategies: 88.3 vs. 90.4 MAP@100. Among sampling strategies, in-batch sampled softmax loss gave the best retrieval results and converged much faster than in-batch cross-entropy, though in-batch triplet loss was fairly similar.

Given that we are using the batch as a source for random negatives, the batch size becomes important. In fact, we found that the larger the batch, the better the retrieval results. Batches of size 2, 10, 100, and 1000 resulted in 82.8, 87.9, 89.2, and 90.4 MAP@100 on Quora.

\section{Conclusion}
\label{sec:conclusion}

In this work, we distinguished between pairwise scoring tasks (including re-ranking) and retrieval tasks. We described a general \textit{dual encoder} abstraction for training arbitrary complex distance functions, and a specific simple setting with negative sampling that improves substantially over standard retrieval baselines.

Our results begin to show that end-to-end retrieval is a viable alternative to discrete retrieval. Future work will include:

\begin{enumerate}
    \item Extending these experiments to larger tasks, with many more retrieval candidates.
    \item Adding a scoring or re-ranking model after retrieval to show overall improvements to existing systems.
    \item Exploiting the Dual Encoder framework presented here to handle multiple data modalities.
\end{enumerate}

%\section*{Acknowledgments}

\newpage
\bibliography{end2endretrieval}
\bibliographystyle{acl_natbib_nourl}

\end{document}